\documentclass[twocolumn]{aastex62}
\usepackage{amsmath}
\usepackage{graphicx}
\usepackage{soul}

\newcommand{\tint}{$T_\mathrm{int}$}
\newcommand{\teq}{$T_\mathrm{eq}$}

\begin{document}
\title{The Intrinsic Temperature and Radiative-Convective Boundary Depth in the Atmospheres of Hot Jupiters}
\author{Daniel Thorngren}
\affil{Department of Physics, University of California, Santa Cruz}
\affil{Institut de Recherche sur les Exoplan\`etes, Universit\'e de Montr\'eal, Canada}
\author{Peter Gao}
\affil{51 Pegasi b Fellow, Department of Astronomy, University of California, Berkeley}
\author{Jonathan J. Fortney}
\affil{Department of Astronomy and Astrophysics, University of California, Santa Cruz}

\begin{abstract}
In giant planet atmosphere modelling, the intrinsic temperature \tint\ and radiative-convective boundary (RCB) are important lower boundary conditions.  Often in one-dimensional radiative-convective models and in three-dimensional general circulation models it is assumed that \tint\ is similar to that of Jupiter itself, around 100 K, which yields a RCB around 1 kbar for hot Jupiters.  In this work, we show that the inflated radii, and hence high specific entropy interiors (8-11 $k_b /$ baryon), of hot Jupiters suggest much higher \tint.  Assuming the effect is primarily due to current heating (rather than delayed cooling), we derive an equilibrium relation between \teq\ and \tint, showing that the latter can take values as high as 700 K.  In response, the RCB moves upward in the atmosphere.  Using one-dimensional radiative-convective atmosphere models, we find RCBs of only a few bars, rather than the kilobar typically supposed.  This much shallower RCB has important implications for the atmospheric structure, vertical and horizontal circulation, interpretation of atmospheric spectra, and the effect of deep cold traps on cloud formation.
\end{abstract}
\keywords{planets and satellites: atmospheres -- planets and satellites: gaseous planets -- planets and satellites: interiors -- planets and satellites: physical evolution}

\section{Introduction}
Soon after the discovery of strongly irradiated giant planets \citep{Mayor1995}, it was realized that they would have strikingly different atmospheres from the giant planets in our own solar system \citep{Guillot1996,Seager1998,Marley1999}.  For Jupiter at optical wavelengths, one can see down to the ammonia cloud tops ($\sim$0.6 bars), which are within the convective region of the planet that extends into its vast deep interior.  For hot Jupiters \citep[with \teq\ $> 1000 K$, see][]{Miller2011}, it was appreciated that their atmospheres could remain radiative to a considerably greater depth due to incident fluxes that are often thousands of times that of Earth's insolation, which force the upper atmosphere to a much higher temperature (typically 1000-2500 K) than for an isolated object. \citep{Guillot2002,Showman2002,Sudarsky2003}.  This leads to a significant departure of the atmospheric temperature structure from an adiabat, and has major consequences for atmospheric circulation \citep{Showman2002,Showman2008, Rauscher2013, Heng2015}. As such, there has long been significant interest in understanding what controls the pressure level of the hot Jupiter radiative-convective boundary (RCB).

Radiative-convective atmosphere models for hot Jupiters found that, for Jupiter-like intrinsic fluxes (parameterized by \tint, of 100 K) but incident stellar fluxes $10^4$ times larger, one typically found RCB pressures near 1 kbar \citep[e.g.,][]{Guillot2002, Sudarsky2003, Fortney2005a}.  While it was understood early on that the RCB depth  strongly depends on the value of \tint\ \citep[e.g.][their Figure 16]{Sudarsky2003}, cooling models suggested that \tint\ values would fall with time to Jupiter-like values \citep{Guillot2002, Burrows2004, Fortney2007}, and the $\sim$~1 kbar RCB became ensconced as a ``typical'' value for these objects.  Such atmospheres are very different from those found in the solar system, so considerable modeling effort has gone into studying their possible vertical and horizontal circulation patterns, and to what degree cold traps at depth may affect what molecules and cloud species can be seen in the visible layers \citep{Hubeny2003, Fortney2008a, Powell2018}.

However, the larger than expected radii of hot Jupiters suggest interiors that are much hotter and more luminous than our own Jupiter \citep[e.g.][]{Guillot2002}.  This is because even a pure H/He object at Jupiter's internal temperatures cannot match the observed radii of hot Jupiters \citep{Miller2011, Fortney2007}.  This implies higher interior fluxes and shallower -- sometimes much shallower -- RCB boundaries than the canonical 1 kbar.  Such atmospheres have occasionally appeared in other works \citep[e.g.][]{Guillot2010, Sing2016, Tremblin2017, Komacek2017}, but not studied extensively.  In this work, we will quantify interior fluxes and the RCB depth as a function of planetary \teq\ to better inform the thermal structure of 1D and 3D atmosphere models.

\section{Modelling}
\begin{figure}[t]
    \centering
    \includegraphics[width=\columnwidth]{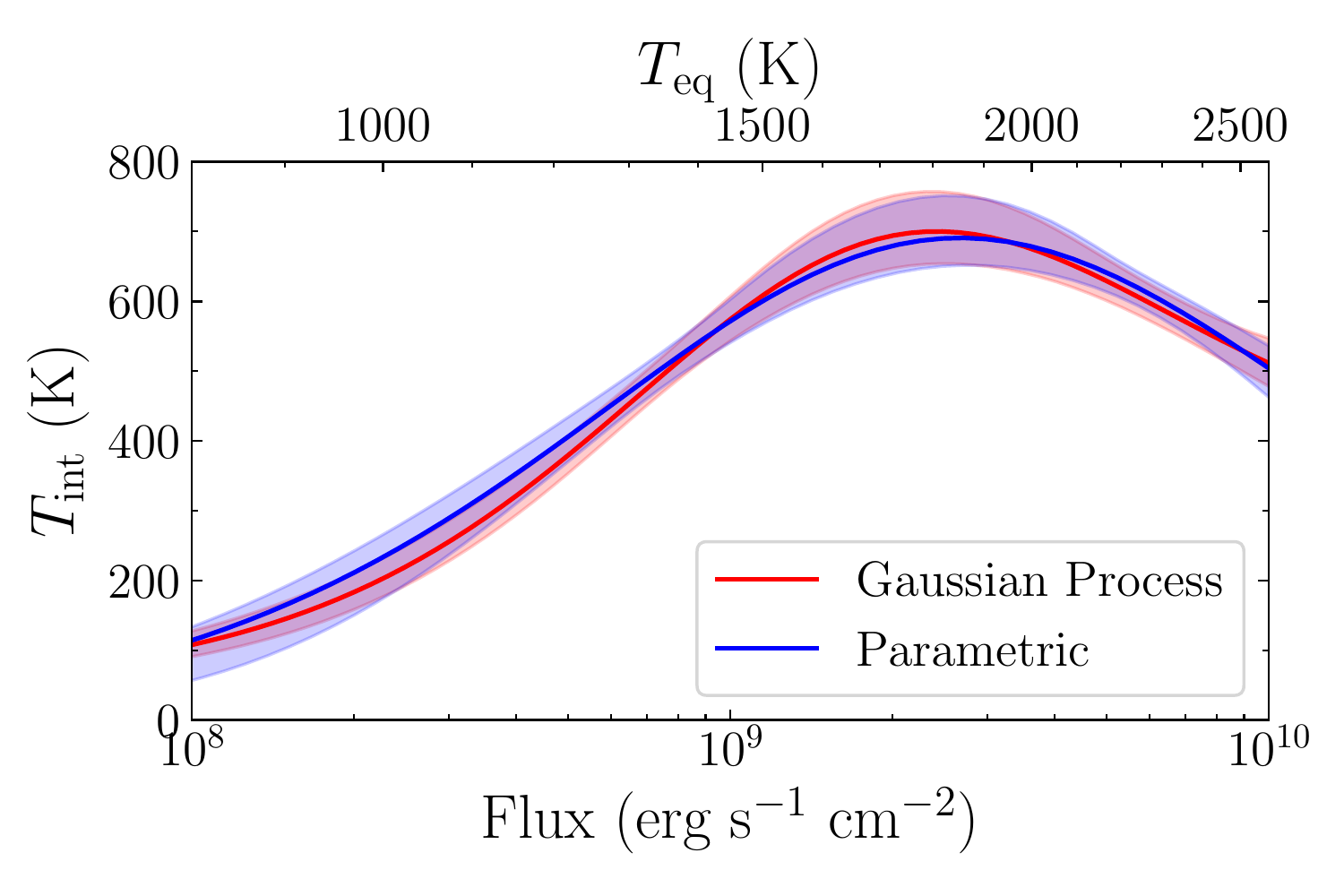}
    \caption{The intrinsic temperatures of hot Jupiters in equilibrium as a function of incident flux (bottom) or equilibrium temperature (top).  These were derived from the two favored heating models (Gaussian process and Gaussian parametric) of \cite{Thorngren2019}, using Eq. \ref{eq:teqToTint}, with corresponding uncertainties.  The two models yield nearly identical results.  Importantly, the intrinsic temperatures must be quite high -- up to 700K -- to match the hot interiors required to explain the radii of hot Jupiters.}
    \label{fig:tint}
\end{figure}

\begin{figure}[t]
    \centering
    \includegraphics[width=\columnwidth]{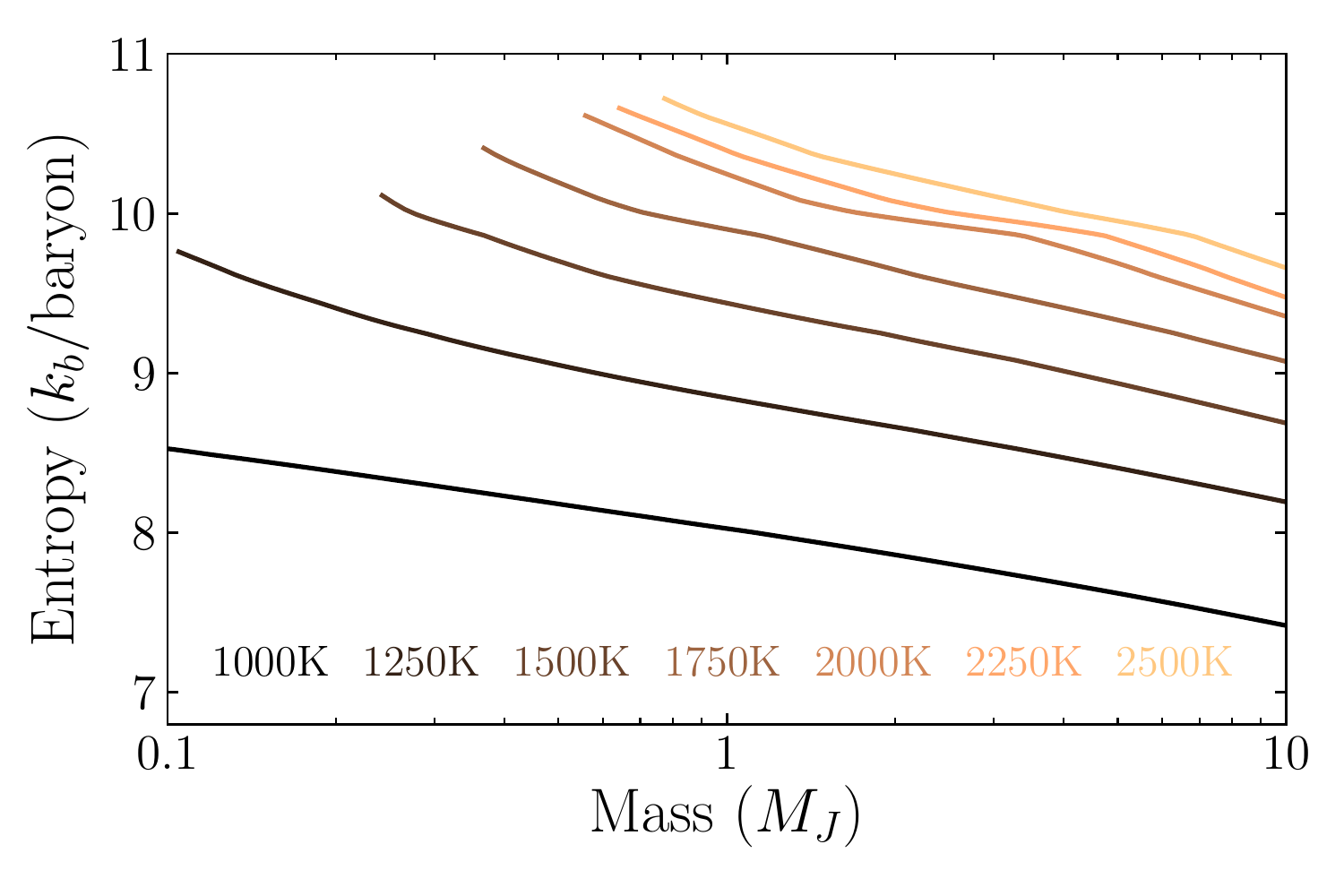}
    \caption{The equilibrium entropy of hot Jupiters as a function of their mass for various equilibrium temperatures.  Each line has models of the same intrinsic and equilibrium temperature, but variations in the resulting surface gravity lead to different internal entropies.  In particular, heat escapes more efficiently through the compact atmospheres of massive objects for a given \tint.  The composition was assumed to be typical \citep[from][]{Thorngren2016}, using the SCvH \citep{Saumon1995} and ANEOS 50-50 rock-ice \citep{Thompson1990} equations of state; different compositions will shift the entropy somewhat.}
    \label{fig:eqEntropy}
\end{figure}

We will parameterize the rate at which heat escapes from a planet's deep interior using the intrinsic temperature \tint.  Its value is primarily driven by the entropy of the underlying adiabat.  Thus, high \tint\ is typical of young exoplanets and inflated hot Jupiters.  If the mechanism inflating hot Jupiters involves the deposition of heat into the interior, then they will eventually reach a thermal equilibrium where $E_\mathrm{in} = E_\mathrm{out}$.  The hotter the planet, the faster this equilibrium will be reached, in as little as tens of megayears \citep{Thorngren2018}.  Once equilibrium is reached, the intrinsic temperature is a function of the equilibrium temperature:
\begin{align}
    4 \pi R^2 \sigma T_\mathrm{int}^4 &= \pi R^2 F \; \epsilon(F)\\
    T_\mathrm{int}  &= \left( \label{eq:teqToTint}
        \frac{F\;\epsilon(F)}{4 \sigma}
    \right)^\frac{1}{4} = \epsilon(F)^\frac{1}{4} T_\mathrm{eq}\\
    & \approx 0.39\; T_\mathrm{eq} \exp\left(
        -\frac{(\log(F)-.14)^2}{1.095}
    \right)
\end{align}
Here, $F$ is the incident flux on the planet (so $F = 4 \sigma T_\mathrm{eq}^4$), $\sigma$ is the Stefan-Boltzmann constant, and $\epsilon$ is the fraction of the flux which heats the interior.  This varies with flux, and was inferred by matching model planets with the observed hot Jupiter population in \citet{Thorngren2018}.  The resulting intrinsic temperatures are shown in Figure \ref{fig:tint}, and the associated entropy (which depend on mass and composition), are shown in Figure \ref{fig:eqEntropy}.

The high intrinsic temperatures this relation produces are important due to the effect they have on the atmosphere.  In particular, the radiative-convective boundary moves to lower pressures for higher \tint.  In contrast, larger \teq\ tends to push the RCB to higher pressures.  As these temperatures are related, it is not immediately obvious where the RCB ends up for planets at high equilibrium temperatures. To evaluate the RCB depth, we generate model atmospheres using a well-established thermal structure model for exoplanets and brown dwarfs \citep[e.g.][]{Mckay1989, Marley1996, Marley1999, Fortney2005a, Fortney2008a, Saumon2008, Morley2012}. The model computes temperature--pressure (TP) and composition profiles assuming radiative--convective--thermochemical equilibrium, taking into account depletion of molecular species due to condensation.

\begin{figure}[t]
    \centering
    \includegraphics[width=\columnwidth]{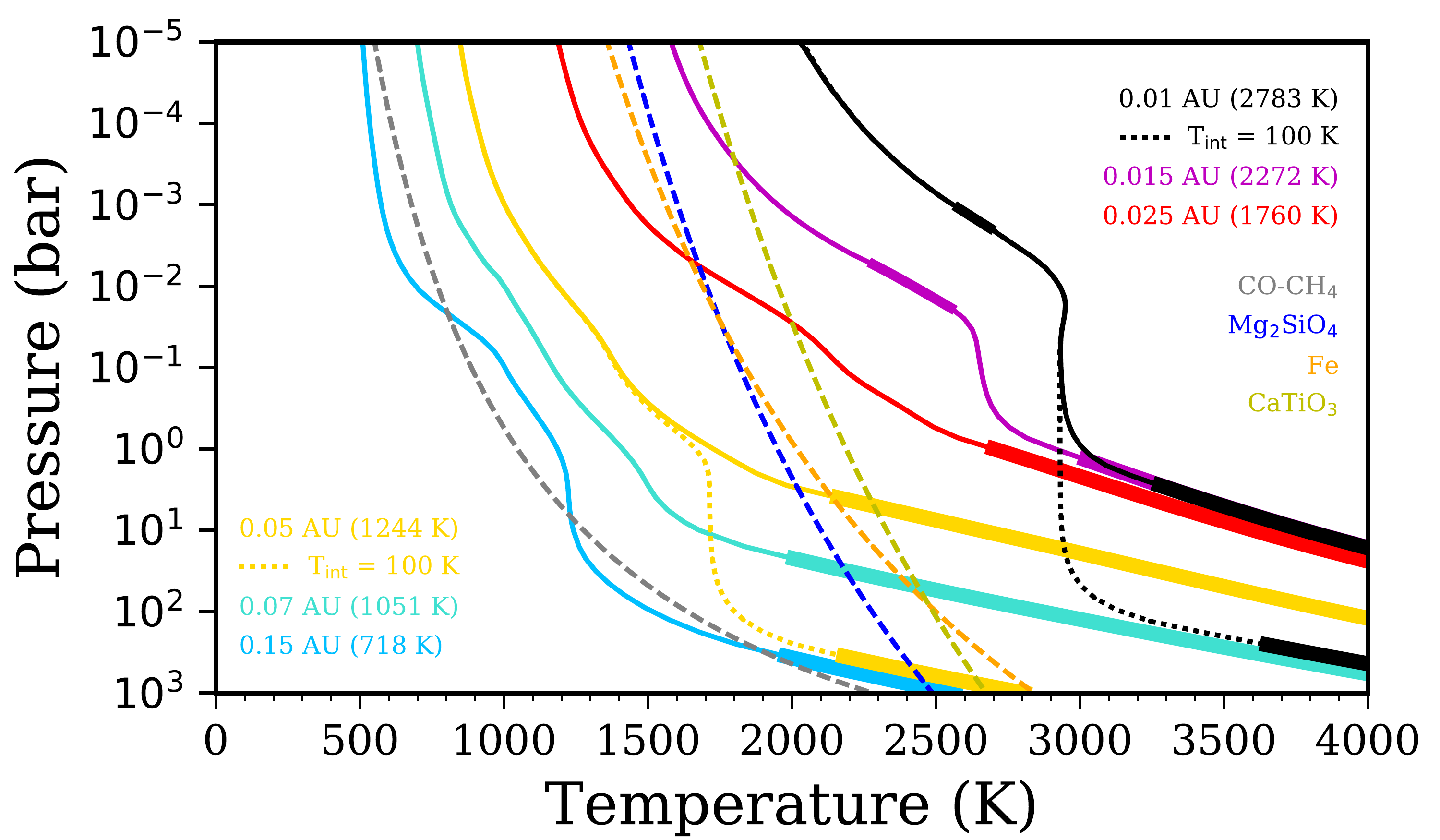}
    \caption{Selected pressure-temperature profiles of our 10$\times$ solar atmospheric metallicity models for various semi-major axes around a sun-like star and the resulting \teq\ (in brackets), from which we derive the intrinsic temperature. The 1-bar gravity is set to 10 m s$^{-2}$.  The thick lines indicate convective regions, whereas thin lines correspond to radiative regions.  Alternative pressure-temperature profiles for a \tint\ = 100 K model for the 0.05 AU and 0.01 AU cases are plotted as dotted curves. The condensation curves for Mg$_2$SiO$_4$, CaTiO$_3$, and iron are shown as dashed curves; \tint\ can be seen to strongly affect their condensation pressures.  The CO-CH$_4$ coexistence curve \citep{Visscher2012} is also shown; in general, hot Jupiters should be well on the CO side.  In the hottest cases, a second convective region forms; however, we will use the term RCB to refer exclusively to the outer edge of the interior adiabatic envelope.  This boundary is visible in the plot for the profiles given, and moves to lower pressures at higher equilibrium temperatures.}
    \label{fig:tp}
\end{figure}

Model atmospheres are generated for a grid of cloud-free giant exoplanets with 1 bar gravities of 4, 10, 25, and 75 m s$^{-2}$ and a range of \teq\ from $\sim$700 to $\sim$2800 K (Figure \ref{fig:tp}).  These were chosen to bracket the gravity and \teq\ of nearly all observed hot Jupiters.  Values of \teq\ were computed assuming full heat redistribution, meaning that incoming stellar radiation is reradiated from the entire planetary surface.  Functionally, we positioned the model planets at various semi-major axes around a sun-like star. Two grids were computed, one assuming solar atmospheric metallicity and one assuming 10 $\times$ solar atmospheric metallicity (similar to Saturn), with any additional heavy elements sequestered in a core, such that the bulk metallicities matched the median of the observed mass--metallicity relationship given by \citet{Thorngren2016}. The RCB depth for each model planet is then defined, when traveling from the deep interior into the atmosphere, as the first pressure level where the local lapse rate transitions from adiabatic to subadiabatic.

We calculate the masses, radii, and adiabat entropies (Figure \ref{fig:eqEntropy}) of our model planets (from the gravity and \teq) using the planetary interior model of \citet{Thorngren2018}, which solves the equations of hydrostatic equilibrium, mass and energy conservation, and an appropriately chosen equation of state (EOS).  We use the SCvH \citep{Saumon1995} EOS for a solar ratio mixture of hydrogen and helium, and the ANEOS 50-50 rock-ice EOS \citep{Thompson1990} for the metals.  

\section{Results and Discussion}
\begin{figure}[tb]
    \centering
    \includegraphics[width=\columnwidth]{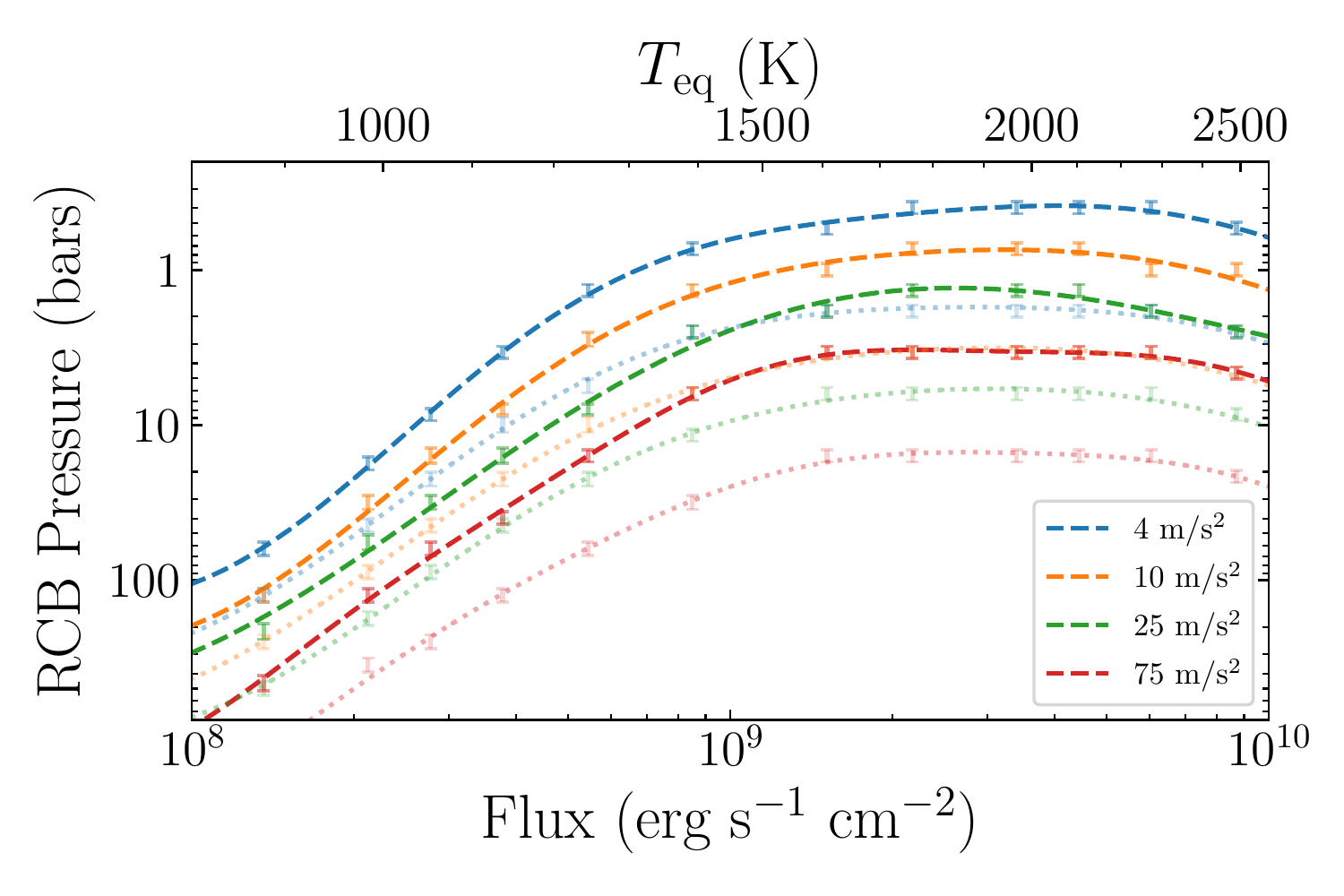}
    \caption{The RCB pressure as a function of incident flux (or \teq), shown for different surface gravities (colors, see legend) and $1\times$ (pale dotted) and $10\times$ (dashed) solar metallicity atmospheres.  Due to binning effects, there are small uncertainties around the modeled points, so we drew smooth lines through the data using Gaussian process interpolation with a squared exponential kernel whose parameters were optimized via the maximum likelihood.}
    \label{rcbLocation}
\end{figure}

Our results for the location of the RCB are shown in Figure \ref{rcbLocation}.  The RCB moves to \emph{lower pressures at higher equilibrium temperatures}, roughly in line with how \tint\ varies with \teq.  At the inflation cutoff of around 1000 K \citep{Miller2011}, the RCB is found at around 100 bars.  At the extremum around $T_\mathrm{eq} = 1800$ K, it is found at roughly 1 bar.  Higher gravity moves the RCB to higher pressures, and higher metallicity moves it to lower pressures. At equilibrium, no hot Jupiter with gravity $<$ 25 m s$^{-2}$ and solar or supersolar atmospheric metallicity should have an RCB as deep as 1 kbar.

\subsection{Relation to the Heating Mechanism}
These results have important implications for understanding the anomalous heating of hot Jupiters.  Heating deposited below the RCB is much more effective for inflating planets than heat deposited above \citep{Komacek2017, Batygin2010}.  This is particularly important for Ohmic dissipation.  For the lower RCB pressures that we predict, models like Ohmic dissipation will be more efficient than previously considered.  Since giant planets are born quite hot, they will have RCBs at low pressures at young ages that could be maintained there by this heating.  However, this would not necessarily allow for reinflation of a planet whose interior has already cooled \citep[see][]{Lopez2016}, as the RCB might already be at tens of bars or deeper when the heating started.  We refer the reader to \cite{Komacek2017} for a more detailed and broader discussion of these heating deposition depth effects.

It is also worthwhile to consider the effect that our assumptions about the hot Jupiter heating have on the model.  In \cite{Thorngren2018}, it was assumed that the heating was proportional to and a function of the incident flux, based on the results of \cite{Weiss2013}.  There may be additional factors that affect the heating, such as planet and stellar mass, but since $\epsilon(F)$ seems to predict planetary radii quite well, these likely add at most modest uncertainty to our estimates of \tint.  An additional consideration is whether the anomalous radii are caused entirely by heating, or whether there is a delayed cooling effect as well; for example, Ohmic dissipation \citep{Batygin2011} may be a combination of these \citep{Wu2013, Ginzburg2016}.  Delayed cooling effects would alter the apparent \tint\ for a given internal adiabat entropy, and delay arrival at thermal equilibrium.  However, many anomalous heating models do not rely on delayed cooling \citep[e.g.][]{Arras2009, Youdin2010, Tremblin2017}, and signs of possible reinflation \citep{Hartman2016, Grunblatt2016, Grunblatt2017} seem to favor these.  If reinflation is conclusively shown to occur, then anomalous heating must be the dominant cause of radius inflation \citep{Lopez2016}, and our \tint\ estimates will be particularly good.  Finally, the usual uncertainties in the equation of state \citep[see e.g.][]{Militzer2013, Chabrier2019} and planet interior structure \citep{Baraffe2008, Leconte2012} discussed in \cite{Thorngren2016} also apply to this work.

\subsection{Effect on Atmospheric Models}
These results have important implications for global circulation models (GCMs) of hot Jupiters.  It has long been a convention in this field to use intrinsic temperatures similar to Jupiter's \citep[e.g.][and many others]{Showman2015, Amundsen2016, Komacek2017a, Lothringer2018, Flowers2019}, around $100$ K \citep{Li2012}.  Our work shows that more realistic values for \tint\ should depend strongly on the incident flux and will typically be several hundreds of Kelvin, as shown in  Figure \ref{fig:tint}.  This difference is important for vertical mixing and circumplanetary circulation, as it shifts the RCB to considerably lower pressures.  It was recently demonstrated in the hot Jupiter context that changing the lower boundary conditions can yield significantly different atmospheric flows in these simulations \citep[see][]{Carone2019}.

The higher implied intrinsic fluxes could also impact interpretations of the observed flux from hot Jupiters.  For phase curves, night-side fluxes will be a mix of intrinsic flux, which in many cases can no longer be thought of as negligible, in addition to energy transported from the day side.  Even on the day side, in near-infrared opacity windows that probe deeply, one might be able to observe this intrinsic flux as a small perturbation on the day-side emission spectrum \citep[e.g.,][]{Fortney2017}

The value of \tint\ is also important for the location and abundance of condensates in hot Jupiter atmospheres. Figure \ref{fig:tp} compares the condensation curves of several species, including forsterite, iron, and perovskite, to our model TP profiles; the intersection between the condensation curve and the TP profile delineates the cloud bases. Previous works that considered low \tint\ atmospheres have hypothesized the existence of deep ``cold traps'' for hot Jupiter clouds, where a cloud base at high pressures ($>$ 100 bars) removes condensates and condensate vapor from the visible layers of the atmosphere \citep[e.g.][]{Spiegel2009,Parmentier2016}. However, higher \tint\ values increase deep atmospheric temperatures, such that the cloud base is much shallower in the atmosphere. For example, at \teq\ = 1244 K, whether \tint\ = 100 K or the nominal value computed in this work can results in differences in the forsterite cloud base pressure of $\sim$2 dex (Figure \ref{fig:tp}). This can have important observable consequences, particular in emission, where the lack of deep cold traps could result in cloudier dayside atmospheres \citep{Powell2018}.

A lack of deep cold traps can also prevent atomic metals in the visible atmosphere from being lost to deep clouds.  While this Letter was under review, \citet{Sing2019} published the detection of singly ionized Mg and Fe in the transmission spectrum of WASP-121b. The fact that these metals are not cold-trapped out of the planet's atmosphere at depth in forsterite and/or iron clouds strongly suggests that the shallow RCB suggested here is correct, at least for this planet (as shown in their Figure 13).

A related effect is the role of the RCB depth in effecting non-equilibrium chemical abundances.  For instance, it is now well-established that CO-CH$_4$ abundances are typically out of equilibrium in cool gas giants and brown dwarfs due to the mixing times being faster than the timescale for CO to convert to CH$_4$ \citep{Cooper2006,Moses2011,Zahnle2014,Drummond2018}.  For the cooler planets modeled in Figure \ref{fig:tp}, if the quench pressure is $\sim$10 to 1000 bars (for instance), then the disequilibrium chemical abundances will differ when the RCB is moved, \citep[see also][]{Drummond2018}, as the local TP profile in the deep atmosphere will move in reference to the local CO-CH$_4$ equal abundance curve.  This effect could be seen in the potential detectability of CH$_4$ in only the very coldest planets modeled here, where the upper atmosphere and deep atmospheres are both relatively cool.

\subsection{Observational Tests and Future Work}

We suggest several approaches to further verify our findings observationally.  The previously supposed \tint\ of 100K could be ruled out if clouds are detected when they would otherwise be cold trapped, particularly for planets with $T_{eq}$ $\sim$ 1100-1600 K \citep[see e.g.][or Figure \ref{fig:tp}]{Lines2018}.  Similarly, low pressure RCBs for hotter objects would lead to the presence of atomic metals or other gaseous species in the upper atmosphere that would otherwise be lost to deep cloud formation \citep{Sing2019}.  In addition, CO-dominated (rather than CH$_4$ dominated) atmospheres across a wide \teq\ range (at least for solar-like C/O ratios), including nearly all models shown in Figure \ref{fig:tp}, would suggest lower pressures RCBs.  Transmission and emission spectroscopy are well-suited to these characterization tasks and the higher precision that will be attained with \emph{JWST} will be important in this area.  More directly, as suggested by \citet{Fortney2017}, high intrinsic fluxes may be measured directly by higher fluxes in the near-IR, particularly in windows in water opacity.  Finally, recent detections of strong magnetic fields suggest that high intrinsic temperatures are the reality \citep{Yadav2017, Cauley2019}, since intrinsic temperature is tied to magnetic field strength \citep{Christensen2009}.  This work now needs to be tied back into revised estimates for the Ohmic dissipation that occurs in these atmospheres.

Future work should focus on the effects that these altered boundary conditions have on the cloud structure, chemical abundances, spectra, and day-night contrasts of hot Jupiters.  As we learn more about hot Jupiter interiors through theoretical developments, population studies (especially from new TESS discoveries) and potentially reinflated giants \citep{Grunblatt2017}, we can better characterize these important atmospheric boundary conditions.

\bibliography{bibliography}
\end{document}